\documentstyle[12pt]{article}
\begin{document}
\begin{titlepage}
\pagestyle{empty}
\title{Geometric interpretation of the 2-index potential as twisted de
Rham cohomology}
\author{TSOU Sheung Tsun\thanks{tsou\,@\,maths.ox.ac.uk} \\
and\\
\and Ioannis P.\ ZOIS\thanks{izois\,@\,maths.ox.ac.uk\,, A.S.\ Onassis
Public Benefit Foundation Hellas Scholar} \\
\\
Mathematical Institute,\\24--29 St.\ Giles', Oxford OX1 3LB, UK.}
\date{}
\maketitle
\begin{abstract}
It is found that the 2-index potential in nonabelian theories does not
behave geometrically as a connection but that, considered
as an element of the second
de Rham cohomology group twisted by a flat connection, it fits well with
all the  properties assigned to it in various physical contexts.  
We also prove some results on the Euler characteristic of the twisted
de Rham complex.
\end{abstract}
\end{titlepage}

\section{The 2-index potential}
A skew rank 2 tensor field arises in various contexts: string theory, 
supergravity, and the loop space formulation of Yang--Mills theory.  For 
notational convenience, we shall consider such a  field 
$B_{\mu\nu} (x)$ interchangeably as a 2-form over spacetime.

In the abelian case, the 2-index field is well studied \cite{hayashi} 
and fits
neatly into the Dirac scheme of fields and potentials for general 
spin \cite{dirac}.  The field $B_{\mu\nu} (x)$ is usually regarded as a
potential transforming under a gauge transformation $\Lambda_\mu (x)$
as
\begin{equation}
\delta B_{\mu\nu} (x) = \partial_\mu \Lambda_\nu (x) - \partial_\nu
\Lambda_\mu (x),   \label{btransf}
\end{equation}
exactly as say the electromagnetic potential but with one more index.
One can also readily define the field strength, as a 3-form
field
\begin{equation}
H_{\nu\rho\sigma}= \partial_\sigma B_{\nu\rho} + \partial_\nu
B_{\rho\sigma} +\partial_\rho B_{\sigma\nu}.
\label{bfield}
\end{equation}

The question immediately arises whether the $B_{\mu\nu}$
field can be interpreted as some sort of connection. 
This point was investigated by
Teitelboim et al \cite{teitel}, and they found that one could regard such
a 2-form as a parallel transport of loops (e.g.\ closed strings),
provided the transformation is abelian, as in (\ref{btransf}).
But for nonabelian $B_{\mu\nu}$ we have to look elsewhere.

Freedman and Townsend \cite{freed}
proposed a Lagrangian for the nonabelian $B_{\mu\nu}$.  Cast in
a first-order formulation of the non-linear $\sigma$ model,
these fields appear as the 
dual of the  Lagrange multiplier giving the flat connecton constraint, thus
\begin{equation}
{\cal L}= {\rm Tr}\, A_\mu A^\mu + {\rm Tr}\, \mbox{}^*\!B_{\mu\nu}
F^{\mu\nu},
\label{siglag}
\end{equation}
where $\mbox{}^*\!B_{\mu\nu}=-\frac{1}{2} \epsilon_{\mu\nu\rho\sigma}
B^{\rho\sigma}$ is the (Hodge) dual of $B_{\mu\nu}$, and Tr denotes the 
trace over the nonabelian indices.  This action is invariant under the
transformation
\begin{equation}
\delta B_{\mu\nu}= D_\mu \Lambda_\nu - D_\nu \Lambda_\mu,
\label{fttransf}
\end{equation}
with $D_\mu=\partial_\mu - ig A_\mu$ the covariant derivative with 
respect to $A_\mu$, but $A_\mu$ itself should not transform.

A similar Lagrangian appears in a loop space formulation of
Yang--Mills theory as a nonlinear $\sigma$ model \cite{poly,cst}:
\begin{equation}
{\cal L} = {\rm Tr}\,{\cal A}_\mu {\cal A}^\mu + {\rm Tr}\, 
\mbox{}^*\!{\cal B}_{\mu\nu} {\cal F}^{\mu\nu},
\label{ymlag}
\end{equation}
where $\cal A$ is the logarithmic derivative of the holonomy of the
gauge potential $A$, and $\cal F$ the covariant curl of $\cal A$.
This is in exact analogy with the Freedman--Townsend Lagrangian
(\ref{siglag}).  Although the loop variables ${\cal A}_\mu$ can be 
thought of as a connection, under a Yang--Mills gauge transformation
(which transforms the potential $A_\mu$ in the usual way), they are 
invariant.  Moreover, the invariance of $\cal L$ under such a
{\em gauge} transformation demands that the $\cal B$ 
field transforms as
\begin{equation}
\delta {\cal B}_{\mu\nu} = {\cal D}_\mu \Lambda_\nu - {\cal D}_\nu 
\Lambda_\mu,
\end{equation}
where ${\cal D}_\mu = \delta_\mu -ig {\cal A}_\mu$ is the loop
covariant derivative corresponding to ${\cal A}_\mu$.  This is 
exactly the Freedman--Townsend transformation (\ref{fttransf}).  
At the same
time, this further confirms the result of \cite{teitel} that
nonabelian $B_{\mu\nu}$ does not behave like a connection, not even in
loop space.

In this paper we shall present a geometric framework in which the
$B_{\mu\nu} (x)$ field is not regarded as a gauge potential but as a
cohomological freedom intimately related to the existence of a flat
connection $A_\mu$.  We go on to explore further mathematical
consequences of this construction which may have useful physical
applications.

\section{Flat connections and the twisted de Rham complex}
For ease of presentation, in this section we shall use almost
exclusively the index-free notation of differential forms.

By a flat connection we mean one with zero curvature.  This means that
we shall include the more general case where the base space $X$ (e.g.\
spacetime) need not be simply connected, in which case a flat
connection may have non-trivial holonomy.  In fact, it is well known
that gauge equivalent classes of flat connections are in 1--1
correspondence with conjugacy classes of irreducible representations
of $\pi_1 (X)$ into the gauge group $G$ \cite{donaldson}.

If we denote the exterior covariant derivative and curvature of the 
connection 
$A$ by $d_A$ and $F_A$ respectively, then on any form $\omega$ we have
\begin{eqnarray}
d^2_A \omega &=& d_A (d_A \omega)   \nonumber \\
 &=& d_A (d \omega + A \wedge \omega)   \nonumber  \\
 &=& d(d \omega + A \wedge \omega) + A \wedge (d \omega + A \wedge
 \omega)  \nonumber  \\
 &=& (dA + A \wedge A) \wedge \omega  \nonumber  \\
&=& F_A \wedge \omega.
\end{eqnarray}
Hence if $A$ is flat, $d^2_A =0$.  This means that the exterior
covariant
derivative can actually be used as the differential in a differential
complex, in direct contrast to the general Yang--Mills case.

Recall that associated to the principal $G$ bundle over $X$, with flat
connection $A$, we have a flat vector bundle $E$ (with fibre the Lie
algebra of $G$).   We can consider the space $\Omega^p (X,E)$ of
$p$-forms with values in $E$, which is by definition the space of
global sections of the vector bundle $(\Lambda^p T^*X) \otimes E$, the
tensor product of the $p$-th exterior power of the cotangent bundle
$T^*X$ and the vector bundle $E$.  Locally over an open
set $U \subset X$ such a $p$-form is given by 
\begin{equation}
\omega = \sum \omega_i \otimes e^i,
\end{equation}
where $\omega_i$ are $p$-forms on $U$ and $e^i$ are sections of $E$
over $U$, and the tensor product is over the algebra of $C^\infty$
functions on $U$.  In our case of a flat vector bundle $E$, we can
extend the usual de Rham complex $\Omega^* (X,d)$ over $X$ to a
complex $\Omega^* (X,E,d_A)$ using the flat connection $A$.  The
flatness guarantees the existence of locally constant sections
$e_U^1, \ldots, d_U^n$ ($n$=rank of $E$) with $d_A e_U^i =0$.  We can
then define the exterior derivative $d_A \omega$ of the form $\omega$
by
\begin{equation}
d_A (\sum \omega_e \otimes e_U^i) = \sum (d\omega_i) \otimes e_U^i
\label{defda}
\end{equation}
over the open set $U$.  Since the sections $e_U^i$ are locally
constant, it can readily be seen that $d_A \omega$ agrees on overlaps
and hence is globally defined \cite{bott}.  Moreover, $d_A^2=0$.
It therefore makes
sense to define the cohomology groups $H^*_A (X,E)$ as $d_A$-closed
forms modulo $d_A$-exact forms in the usual way.  It is easy to see
that if $E$ is a trivial bundle of rank $n$ with the trivial flat
connection, then $H^*_{\rm trivial}
(X,E)$ is just $n$ copies of the usual de Rham groups $H^*(X)$.

It is generally recognized that cohomology group elements
correspond to physically interesting quantities \cite{witten}.  
If we now think of
the $B$ field not as a 2-from but as an element of $H^2$, then its
transformation is nothing but the cohomological
freedom of an exact 2-form:
\begin{equation}
\delta B= d_A \Lambda
\label{formtransf}
\end{equation} 
with $\Lambda$ a 1-form, in other words, the transformation 
(\ref{fttransf}).  It is therefore {\em not} a gauge freedom of
the usual Yang--Mills type.  Moreover, (\ref{formtransf}) reduces to
(\ref{btransf}) in the abelian case, which need not therefore be
interpreted as a gauge (in the electromagnetism sense)
transformation.  In addition, the 3-form $d_A B$ is as that
discussed \cite{thierry} for the `curvature' of $B$.

As emphasized in \cite{bott} and obvious from the definition
(\ref{defda}), the cohomological groups depend in
general on the particular trivialization chosen for $E$.  This means
that, if we think of $A$ as a connection in a principal bundle (as in
the gauge case), then gauge equivalent $A$'s may give rise to different
$B$'s.  This makes perfect sense for the theory in hand, because the
term Tr$\,A^2$ in (\ref{siglag}) makes it immediately obvious that the
Lagrangian $\cal L$ is {\em not} `gauge invariant'.  This is why
whereas $B$ transforms as in (\ref{fttransf}), $A$ must remain
invariant.

The same observations apply to the loop space formulation of
Yang--Mills theory.  Since the phase factor is
Yang--Mills gauge invariant, the loop space connection $\cal
A$ is also gauge invariant.  So is of course the Lagrangian in
(\ref{ymlag}).  On the other hand, there is {\rm no} freedom in
transforming the loop connection $\cal A$, because that would mean
changing the phase factor which is a physically measurable quantity.
  
The twisted de Rham cohomology groups $H^*_A (X,E)$ are topological
invariants which are defined whenever there is a flat connection on a
vector bundle $E$.  Now a flat connection appears in many contexts
which may be physically interesting, notably in integrable systems.
This is not surprising: a flat connection ensures integrability of
lifts.  The following results are easy consequences of the
definitions and may prove useful in studying the invariances of such
systems.

In analogy with the usual Euler characteristic of a manifold, we make
the following definition\footnote{Applying noncommutative geometry 
methods to the
flat foliation induced by the flat connection $A$, one can study 
the $\eta$-invariant (related to global anomalies)
which is more sensitive than the
Euler characteristic defined here 
but less sensitive than the de Rham cohomology
groups: it depends on the gauge equivalence class of $A$.  
This will be reported elsewhere \cite{zois}.}.

\vspace{3mm}

{\noindent}{\bf Definition} The {\em Euler characteristic} of the
twisted de Rham complex \newline
$\Omega^* (X,E,d_A)$ is defined to be
\begin{equation}
\chi (X,E) = \sum_i (-1)^i \dim (H_A^i (X,E)).
\end{equation}

The notation makes sense because of the following result.

\vspace{3mm}

{\noindent}{\bf Proposition} {\em Let} $E$ {\em be the 
adjoint vector bundle}
ad$P${\em , associated to the principal bundle} $P$ {\em over a
manifold} $X$ {\em with structure group} $G$ {\em assumed compact and
connected, equipped with a flat connection} $A$.  {\em With respect to
the induced connection} $E$ {\em is a flat vector bundle.  Then the
Euler characteristic} $\chi (X,E)$ {\em is independent of the flat
connection used in calculating the cohomology groups} $H_A^i (X,E)$.

\vspace{3mm}

{\noindent}{\bf Proof} \   We shall prove this by calculating the Euler
characteristic $\chi (X,E)$ using the symbol of an elliptic operator
associated to the differential $d_A$.

For simplicity we shall use the same symbol $d_A$ for all the
differentials in the complex:
\begin{equation}
(d_A)_p \colon (\Lambda^p T^*X) \otimes E \to (\Lambda^{p+1} T^*X) 
\otimes E.
\end{equation}
We can `assemble' the bundle (for details see \cite{atiyah}) 
by defining the single operator
\begin{equation}
D_{d_A} \colon \Omega^{\rm even} \to \Omega^{\rm odd},
\end{equation}
where 
\begin{eqnarray*}
\Omega^{\rm even}&=& \Gamma (\bigoplus_p (\Lambda^{2p} T^*X) \otimes
E),\\
\Omega^{\rm odd}&=& \Gamma (\bigoplus_p (\Lambda^{2p+1} T^*X) \otimes
E),
\end{eqnarray*}
defined by
\begin{equation}
D_{d_A} = d_A \oplus d_A^*,
\end{equation}
that is,
\begin{equation}
D_{d_A} (\omega_0, \omega_2, \ldots) = (d_A \omega_0 + d_A^* \omega_2,
d_A \omega_2 + d_A^* \omega_4, \ldots),
\end{equation}
where $d_A^*$ is the formal adjoint of $d_A$ with respect to some
Riemannian metric on $X$.

Recall the symbol $\sigma (D)$ of a differential operator from
sections of a vector bundle $E$ to sections of a vector bundle $F$
\begin{equation}
D \colon \Gamma (E) \to \Gamma (F)
\end{equation}
is a vector
bundle homomorphism
\begin{equation}
\sigma (D) \colon \pi^* (E) \to \pi^* (F),
\end{equation}
where, for $SX$ the unit sphere bundle in the tangent bundle, $\pi$ is
the canonical projection $SX \to X$.  In local coordinates, 
since $D$ is first order in this case, $\sigma
(D)$ is obtained by replacing $\partial /\partial_\mu$ with $i
\xi_\mu$, where $\xi_\mu$ is the $\mu$th coordinate in the cotangent
bundle $T^*X$.  Furthermore, $D$ is elliptic if its symbol is
invertible.  This can be extended to a differential complex $E$ which
is elliptic if the corresponding sequence of symbols $\sigma (E)$ is
exact outside the zero section of $TX$.

For the flat bundle $E$, if we denote by $\Delta_p$ the Laplacian on
$\Omega^p (X,E)$, thus
\begin{equation}
\Delta_p = (d_A)_{p-1} (d^*_A)_{p-1} + (d_A^*)_p (d_A)_p,
\end{equation}
then (since $d_A^2 = {d_A^*}^2 =0$)
\begin{eqnarray*}
D_A D_A^* &=& \bigoplus \Delta_{2p} \\
D_A^* D_A &=& \bigoplus \Delta_{2p+1}.
\end{eqnarray*}

The exactness of the symbol complex $\sigma (\Omega (X,E))$ off the
zero section then implies that $\sigma (\Delta_p)$ is an isomorphism
(off the zero section). Therefore, $\Delta_p$ and hence $D_A$ are
elliptic.  Then it follows from the usual Hodge theory that
\begin{eqnarray*}
{\rm Ker} D_A & = & \bigoplus h^{2i} \\
{\rm Coker} D_A &= & \bigoplus h^{2i+1}
\end{eqnarray*}
where $h^i$ are the harmonic sections of the bundle $\bigoplus (\Lambda^p
T^*X) \otimes E$, namely elements of Ker\,$\Delta^i$.  This means we
have found an elliptic operator $D_A$ whose index gives the required
Euler characteristic:
\begin{equation}
{\rm ind} (D_A) = \chi (X,E).
\end{equation}
By the Atiyah--Singer index formula \cite{atiyah}, we know that this
depends only on the symbol of $D_A$ and not on $D_A$ itself.

It is obvious from the above that the symbol of $d_A$ is independent
of the flat connection $A$ used, since the term of highest degree is
$\partial /\partial_\mu$.  The symbol of
$D_A$ is given by $i\xi - i\xi^*$ (where $\xi^*$ is contraction with
$\xi$), also independent of $A$.  The index of $D_A$ then gives the
Euler characteristic as above, which is therefore
independent of the flat connection $A$
used.  \hfill $\Box$

\vspace{3mm}

{\noindent}{\bf Corollary 1} {\em When} $X={\bf R}^4,\ \chi(X,E)=$dim\,$G$.

{\noindent}{\bf Proof}\  Obvious. \hfill $\Box$

\vspace{3mm}

{\noindent}{\bf Corollary 2} {\em Under a usual gauge transformation,   
the Freedman--Townsend theory remains in the same
stable isomorphism class (of complex vector bundles over} $X$).

{\noindent}{\bf Proof} \  The
homotopy classes of symbols of elliptic pseudo-differential 
operators are in
one-one correspondence \cite{atiyah} with elements of $K^0 (TX)$,
where $TX$ is the tangent bundle of $X$, which in turn can be
identified with the analytic $K$-homology group $K_0 (X)$ \cite{baum}. 
The former
is by definition stable isomorphism classes of vector bundles over $X$. 
\hfill $\Box$

\section{Remarks}
In the case when $B_{\mu\nu}$ is abelian, it can be shown easily
\cite{freed} that the theory is equivalent to a massless scalar field.
This is an example of the general duality between scalar fields and
($d-2$)-form fields, where $d$ is the dimension of spacetime
\cite{hitchin}.  Here $d=4$.
This duality also interchanges Bianchi identities (topology) and
equations of motion (dynamics), reminiscent of the Wu--Yang treatment
of electric and magnetic charges \cite{wuyang,cst}.  Similar considerations
apply in the nonabelian case, giving the equivalence between the
first-order and second-order formulations of the non-linear $\sigma$
model \cite{freed}.  Here the scalar field is obtained from the flatness
condition of $A_\mu$, which is locally of the form $g^{-1}\partial_\mu
g$, with $g$ an element of the group $G$.

One may ask where the extra degrees of freedom of a spin 2 field have
gone to, if it is equivalent to a scalar field.  This is exactly
accounted for by its cohomological freedom (\ref{formtransf}).  
Suppressing the Lie algebra indices, the 6
degrees of freedom of a skew rank 2 tensor are taken up by the 4
degrees of the vector $\Lambda_\mu$, plus {\em its} cohomological
freedom of an additive scalar, leaving just the one degree of freedom
of a scalar field.

In the case of Yang--Mills theory in loop space, this extra freedom
gives rise to a {\em dual} gauge symmetry which is magnetic in nature
if the original symmetry is considered to be electric.  This leads to
a fascinating electric--magnetic dual symmetry for Yang--Mills theory
which is somewhat unexpected \cite{cft}.

In conclusion, the interpretation of the 2-index field as a twisted de
Rham cohomology group element, together with its inherent cohomological
freedom, gives a geometric explanation of many of its properties.  This
is particularly interesting for the nonabelian case and may serve as
a guide for studying its possible interactions.  Furthermore, this
geometric interpretation gives a satisfying picture for the loop space
formulation of Yang--Mills theory, with particular regard to its
symmetry properties.

\end{document}